\documentclass[twocolumn]{aastex7}

\begin{document}

\title{Investigating the Anti-Correlation between Photon Index and Flux of the Crab using {\it RXTE} and {\it NuSTAR}}

\correspondingauthor{Debjit Chatterjee}
\email{debjitchatterjee92@gmail.com, debjit@mx.nthu.edu.tw}
\correspondingauthor{Hsiang-Kuang Chang}
\email{hkchang@mx.nthu.edu.tw}

\author[0000-0001-6770-8351]{Debjit Chatterjee}
\affiliation{Institute of Astronomy, National Tsing Hua University, Hsinchu 300044, Taiwan}
\email{debjitchatterjee92@gmail.com, debjit@mx.nthu.edu.tw}

\author[0000-0002-5617-3117]{Hsiang-Kuang Chang}
\affiliation{Institute of Astronomy, National Tsing Hua University, Hsinchu 300044, Taiwan}
\affiliation{Department of Physics, National Tsing Hua University, Hsinchu 300044, Taiwan}
\affiliation{Institute of Space Engineering, National Tsing Hua University, Hsinchu 300044, Taiwan}
\affiliation{Center for Theory-Computation-Data Science Research (CTCD), National Tsing Hua University, Hsinchu 300044, Taiwan}
\email{hkchang@mx.nthu.edu.tw}

\author[0000-0003-1856-5504]{Dipak Debnath}
\affiliation{Institute of Astronomy, National Tsing Hua University, Hsinchu 300044, Taiwan}
\email{dipakcsp@gmail.com}

\author[0009-0008-7888-7584]{Koothodil Abhijith Augustine}
\affiliation{Institute of Astronomy, National Tsing Hua University, Hsinchu 300044, Taiwan}
\email{abhijithaugustine007@gmail.com}

\author[0009-0006-0904-8301]{Tzu-Hsuan Lin}
\affiliation{Department of Physics, National Tsing Hua University, Hsinchu 300044, Taiwan}
\email{qtp010629@gmail.com}

\begin{abstract}
We present a systematic study of the recently reported anti‑correlation between X‑ray flux and photon index ($\Gamma$) in the Crab Nebula, using archival RXTE/PCA (3 -- 50 keV), RXTE/HEXTE (20 -- 100 keV), and NuSTAR (3 -- 78 keV) observations. Spectra were extracted in soft (3 -- 10 keV) and hard bands (10 -- 50 keV, 10 -- 78 keV, 20 -- 100 keV) and fitted with an absorbed power‑law model. Across all instruments and energy ranges, we confirm the existence of a persistent negative correlation -- harder spectra at higher flux levels. The correlation is stronger in the hard bands compared to the soft bands. This is consistent with synchrotron emission modulated by magnetic‑field variations in the pulsar wind nebula.
\end{abstract}



\section{Introduction} \label{sec:intro}

Pulsar wind nebulae (PWNe) are dynamic, high-energy astrophysical systems formed when a rapidly rotating neutron star injects a relativistic wind of particles into the surrounding medium following a supernova (SN) explosion. The rotational energy loss of the pulsar powers this wind, which inflates a magnetized bubble of relativistic particles inside the supernova remnant (SNR). This structure, known as a PWN, emits across the electromagnetic spectrum and is often detected as a source of non-thermal radiation. The dominant emission mechanisms in PWNe are synchrotron radiation and inverse Compton scattering. Relativistic electrons spiraling along the magnetic field lines emit synchrotron radiation from radio to hard X-ray bands, while the same electrons can up-scatter ambient low-energy photons to gamma-ray energies via inverse Compton processes \citep{Slane2017}. Observations suggest that PWNe contribute significantly to the population of Galactic very-high-energy (VHE) leptonic sources.

However, despite their prominence, key questions regarding particle acceleration and transport in PWNe remain unresolved. While diffusive shock acceleration (DSA) has generally been proposed to explain high-energy particle production, it struggles to account for complex features such as the multiple spectral breaks observed in the Crab Nebula, from radio wavelengths to soft gamma-rays \citep{Atoyan1999, Bietenholz2001, Dubner2017, Tanaka2017}. Moreover, the observed morphological differences between low-energy and high-energy structures within the nebula challenge the predictions of simple DSA models. An alternative explanation involves magnetic reconnection in the striped structure of the pulsar wind, which may efficiently transfer magnetic energy into particle acceleration near the wind termination shock \citep{Sironi2011}. This scenario predicts a broken power-law distribution of electron energies, in contrast to the simpler power-law expected from Fermi acceleration, and naturally explains the required change in magnetization from the vicinity of the pulsar to the wind termination shock region.

The Crab Pulsar (PSR B0531+21), a rapidly rotating neutron star with an intense magnetic field, is a key component of the Crab nebula \citep{LyneGrahamSmith2012}. This pulsar injects energy into the nebula, driving synchrotron radiation \citep{Shklovsky1968}. Understanding the pulsar-nebula interaction is crucial for comprehending PWN dynamics and particle acceleration in astrophysical environments \citep{KennelCoroniti1984}.
Despite its long-standing reputation as a standard candle in high-energy astrophysics, variability in the Crab’s emission has been detected. Unexpected gamma-ray flares, observed by the Fermi Gamma-ray Space Telescope and AGILE, suggest dynamic magnetic reconnection events within the nebula, challenging conventional models of pulsar wind nebulae \citep{Buhler&Blandford2014, Reynolds2016}. Additionally, secular trends in the X-ray and gamma-ray fluxes of the pulsar have been linked to its spin-down evolution, offering a unique probe into pulsar magnetospheric physics \citep{Yan2018, Zhao2023}.
Recent observations reveal complex structures within the Crab Nebula, including jets and filaments shaped by the pulsar wind and supernova ejecta \citep{Hester2008}. These features offer clues about the mechanisms of energy transportation and particle acceleration \citep{Tanaka2011}. Gamma-ray flares detected in the Crab Nebula challenge existing models and prompt new research into particle acceleration and emission \citep{Tavani2011}.

Disentangling these scenarios requires detailed spectral and spatial analysis across a broad energy range. While synchrotron cooling can also produce spectral steepening, this effect must be separated from injection-related features. X-ray spectral softening observed downstream of the termination shock in many PWNe \citep{Mori2004, Kargaltsev2008, Ma2016, Liu2023} is generally attributed to synchrotron cooling, but does not rule out more complex acceleration dynamics.

Previous studies have demonstrated a significant anti-correlation between X-ray flux and photon index of the Crab \citep{Augustine&Chang2024} using {\it Swift}/BAT data in the 15--150 keV energy band to identify a statistically significant anti-correlation between the Crab Nebula’s hard X-ray flux and its photon index. This result suggested that variations in magnetic field strength within the nebula influence the acceleration of relativistic electrons by DSA, leading to observable spectral changes. While that study provided compelling evidence for synchrotron processes governed by magnetic variability, it was limited by energy range, sparse temporal sampling, and a single instrument. Motivated by these findings, the present study aims to investigate the photon index–flux anti-correlation using a broader and more continuous dataset, spanning 3--100 keV, obtained from {\it RXTE}/PCA, {\it RXTE}/HEXTE, and {\it NuSTAR}. By incorporating data from multiple instruments and covering both soft and hard X-ray regimes, this work offers a more comprehensive and statistically robust validation of the anti-correlation. It also explores whether the phenomenon persists across different energy bands and observational epochs, providing strong evidence of the anti-correlation properties.

The paper is organized in the following way: in Section~\ref{sec:obs}, we discuss the observation and data reduction procedure. Section~\ref{sec:result} demonstrates the spectral analysis method and the results obtained. Next, we discuss the results and compare them with previous studies in Section~\ref{sec:discus}.

\section{Observation and Data Reduction}\label{sec:obs}
We have processed the data using HeaSoft version 6.34 and the FTOOLs\footnote{https://heasarc.gsfc.nasa.gov/ftools} package.

\subsection{RXTE}
The Proportional Counter Array (PCA) onboard the {\tt Rossi X-ray Timing Explorer} ({\it RXTE}) was designed to provide high-time-resolution and broadband X-ray spectral coverage in the 2--60 keV energy range \citep{Jahoda1996}. The High Energy X-ray Timing Experiment (HEXTE) onboard the {\it RXTE} provided spectral coverage in the 15--250 keV energy range, also with high time resolution \citep{Rothschild1998}. These two instruments make {\it RXTE} well suited to studying the spectral and temporal properties of X-ray sources.

We used the archival data of the PCA \citep{Jahoda1996} 
and HEXTE \citep{Rothschild1998} to study the Crab. The PCA data are selected from April 8, 1996 (MJD = 50181.62) to December 31, 2011 (MJD = 55926.99). The HEXTE observations are selected from January 7, 1996 (MJD=50089.10) to November 18, 2005 (MJD=53692.75). We excluded the observations with low exposure times for better statistics during spectral fits. We only selected the targeted observations, excluding the slew and scan mode observations. 

We followed the standard procedure to extract the PCA and HEXTE spectra described in `The ABC of XTE'\footnote{https://heasarc.gsfc.nasa.gov/docs/xte/abc/front\_page.html}. For PCA, we used data from Proportional Counter Unit 2 (PCU2) in Standard-2 mode (FS4a*). We generated a background fits file using the {\tt PCABACKEST} tool and the `bright' background model appropriate for our observation periods. A good time interval (GTI) file is created using the FTOOLS task {\tt maketime} to include only periods when the instrument operates under optimal conditions. The {\tt saextrct} tool was used to extract the source spectra and background spectra using the GTI file. The spectra were generated in two energy bands 3 -- 10~keV and 10 -- 50~keV. To improve the signal-to-noise ratio, we rebinned the spectra using the {\tt rbnpha} tool, combining adjacent energy channels to ensure each bin had a minimum number of counts. The spectra were rebinned to have at least 5 counts/bin to obtain valid $\chi^2$ statistics. We then generated the response matrix and effective area files using the {\tt pcarsp} tool to account for the instrumental response.

We used HEXTE science mode data (FS52* in the energy range 20 -- 100 keV) of Cluster 0 or A. The first step involves generating the background spectra using the {\tt hxtback} tool, which creates the source and background files by filtering event data based on cluster positions.
The extraction of the source and background spectra is performed using the {\tt saextrct} tool. The GTI file is applied to ensure valid time intervals. Then instrumental dead-time correction is applied using the {\tt hxtdead} tool, which modifies both source and background spectra to account for dead-time effects, ensuring accurate exposure times. The task {\tt hxtrsp} is to generate response matrix file (RMF). 

\subsection{NuSTAR}
{\tt Nuclear Spectroscopic Telescope Array} ({\it NuSTAR}) is a space-based X-ray telescope launched by NASA on June 13, 2012. It is the first telescope capable of producing focused images of high-energy X-rays in the range of 3 to 79 keV. {\it NuSTAR} telescope comprises two identical focal plane modules, FPMA and FPMB \citep{Harrison2013}. {\it NuSTAR} observed the Crab multiple times from 2012 till now as part of the instrument calibration campaign. We excluded the observations with low exposure (3~ks) time (which eventually results in bad spectral fits). We used a total of 61 archival observations from 2012 September 20 to 2024 March 18 for our analysis.

Although Crab is a bright source ($\sim$250 cts/s  per FPM with a $\sim$50\% dead time), no special processing or pile-up corrections are required \citep{Madsen2015}. This is due to the focal plane detectors' triggered readout system, which features a short preamplifier shaping time of 1~$\mu$s \citep{Harrison2013}. The raw data were reprocessed using the {\it NuSTAR Data Analysis Software} ({\tt NuSTARDAS} version 2.1.2). Calibrated and cleaned event files were generated with the {\tt nupipeline} task, utilizing the 20200912 version of calibration files from the {\it NuSTAR} calibration database\footnote{http://heasarc.gsfc.nasa.gov/FTP/caldb/data/nustar/fpm/}.

Source and background spectra were extracted using circular regions with a $60''$ radius. The background region was chosen far from the source region with $90''$ radius. Spectra were obtained from the cleaned science mode event files using the {\tt nuproducts} task. We used both the FPMA and FPMB spectra for our analysis. We generated spectra in two energy bands: 3 -- 10 keV (soft band), 10 -- 78 keV (hard band). The spectra were re-binned to a minimum of 25 counts per bin using the {\tt grppha} task to have a good fit statistics.

\begin{figure*}[ht!]
\plotone{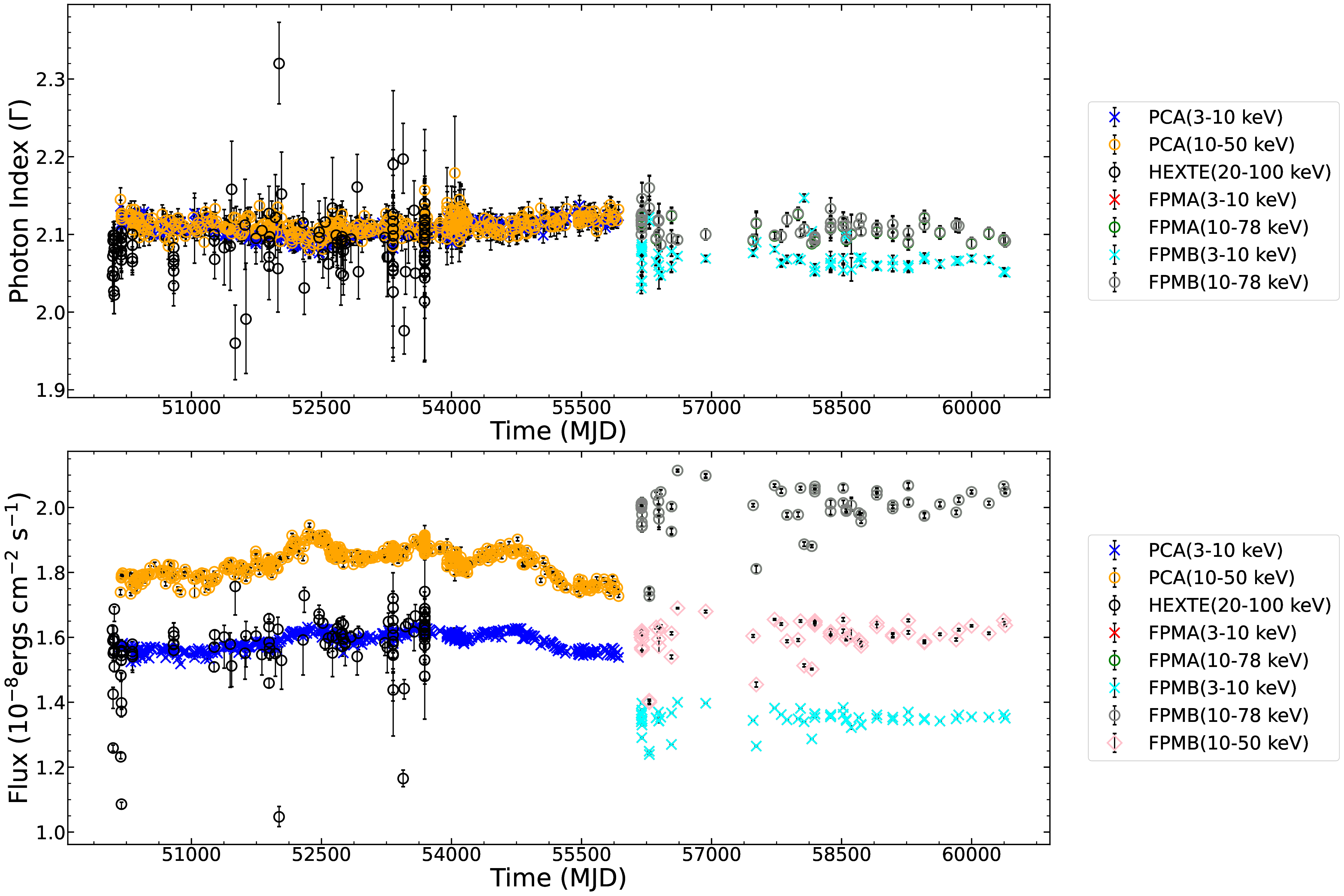}
\caption{Variation of photon index (upper panel) and flux (lower panel) with MJD for PCA (3--10 keV; blue), PCA (10--50 keV; yellow), HEXTE (20--100 keV; black), FPMA (3--10 keV; red), FPMA (10--78 keV; green), FPMB (3--10 keV; cyan), FPMB (10--78 keV; gray), FPMB (10--50 keV; pink). The errors are estimated with 90\% confidence.  
{\tt Note:} The similar results of FPMA and FPMB are almost overlapped in the figure. We have only plotted the 10--50~keV FPMB flux, as the values are similar to FPMA (10--50~keV)\label{fig:all_variation}}
\end{figure*}

\section{Results}\label{sec:result}
We use HeaSoft's spectral analysis package {\tt XSPEC}\footnote{https://heasarc.gsfc.nasa.gov/xanadu/xspec/}\citep{Arnaud1996} version 12.14.1 to fit the extracted spectra of {\it RXTE}/PCA (3--10 keV and 10--50 keV), {\it RXTE}/HEXTE (20--100 keV) and {\it NuSTAR}/ FPMA and FPMB (3--10 keV and 10--78 keV). We used $\chi^2$ statistics to justify the goodness of the spectral fit. All the errors are estimated with 90\% confidence throughout the manuscript unless otherwise quoted.

\subsection{Spectral Analysis}
We fitted the spectra using a power-law model (PL). To incorporate the galactic absorption we used {\tt Tbabs} model. The $\rm N_{\rm H}$ value is fixed to $3.6\times10^{21}~\rm cm^{-2}$\citep{Ge2012}. We use \textsc{vern} scattering cross-section \citep{Verner1996} and \textsc{wilm} abundances \citep{Wilms2000}.

\subsection{RXTE}
Among all the good observations of PCA, a total of 437 observations fit well (in both 3--10 keV and 10--50 keV energy range) with our combined model: \textsc{Tbabs*PL}. We have excluded the observations with bad fits based on the obtained $\chi^2_{\rm red}$ values. The observations given a $\chi^2_{\rm red}$ values within the 90\% confidence range are considered for our final study. Similar criteria are also applied for the HXTE observations, and we obtained 120 HEXTE best-fit spectra. During our study we noticed the variation of the unabsorbed PL-flux for both the 3--10 keV (blue crosses in the lower panel of Fig.~\ref{fig:all_variation}) and 10--50 keV (yellow circles in the lower panel of Fig.~\ref{fig:all_variation}) energy ranges for PCA spectra, as well as for the 20--100~keV energy band for HEXTE spectra (black circles in the lower panel of Fig.~\ref{fig:all_variation}). The variation of the photon index ($\Gamma$) is not very significant and remains $\sim$2.1 for all the energy bands. It is also noticed that the $\Gamma$ values are marginally higher in the 10--50~keV (yellow crosses in the upper panel of Fig.~\ref{fig:all_variation}) range than that for the 3--10~keV (blue crosses in the upper panel of Fig.~\ref{fig:all_variation}). The flux values are always lower in the 3--10~keV range than the 10--50~keV. However, the 20--100-keV flux shows similar values as the 3--10~keV flux along with some random fluctuations. 

\subsection{NuSTAR}
The extracted FPMA and FPMB spectra are also fitted separately with the same combination of models. The $\Gamma$ for both 3--10~keV (red and cyan crosses in the upper panel of Fig.~\ref{fig:all_variation}) and 10--78~keV (green and gray circles in the upper panel of Fig.~\ref{fig:all_variation}) energy ranges show a slight variation during our observation periods. The $\Gamma$ values for different energy bands are significantly different for both the FPMA and FPMB spectra. Similar to the PCA observations, the 3--10~keV band $\Gamma$ values are lower than the 10--78~keV band values except for some fluctuations. The flux values also show similar trends as of PCA spectra --  3--10~keV flux (red and cyan crosses in the lower panel of Fig.~\ref{fig:all_variation}) are lower than the 10--78~keV flux (green circles in the lower panel of Fig.~\ref{fig:all_variation}). In general, we notice similar results for FPMA and FPMB spectral fits. Our findings are in agreement with the results based on 2015 and 2016 {\it NuSTAR} observations by \citet{Madsen2015}.

\begin{deluxetable*}{l|ccccccch}
\tablecaption{Pearson and Spearman Correlation Results\label{tab:corr}}
\tablewidth{0pt}
\tablehead{
\colhead{Correlation} & \multicolumn{3}{|c|}{{\it RXTE}} & \multicolumn{4}{c}{{\it NuSTAR}} & \nocolhead{Swift}  \\
\colhead{} & \multicolumn{2}{|c}{PCA} &  \colhead{HEXTE}&  \multicolumn{2}{|c}{FPMA}&  \multicolumn{2}{c}{FPMB} &  \nocolhead{XRT} \\
\colhead{} & \colhead{3-10 keV} & \colhead{10-50 keV} &  \colhead{20-100 keV}&  \colhead{3-10 keV} & \colhead{10-78 keV}&  \colhead{3-10 keV} & \colhead{10-78 keV} & \nocolhead{3-10 keV} 
}
\decimalcolnumbers
\startdata
Pearson & -0.48 & -0.53 & -0.52 & -0.33  & -0.57 & -0.33 & -0.56 & -0.34   \\
p value & $10^{-27}$ & $10^{-33}$ & $10^{-9}$ & 0.01 & $10^{-6}$ & $ 0.01 $  & $10^{-6}$ & 0.07\\
\hline
Spearman & -0.48 & -0.56 & -0.42 & 0.04 & -0.51 & 0.04 & -0.51 & -0.33  \\
p value & $10^{-24}$ & $10^{-37}$ & $10^{-6}$ & 0.77 & $10^{-5}$ & 0.77 & $10^{-5}$ &  0.08 \\
\enddata
\tablecomments{The correlation coefficients and their respective p-values for different instruments and in different energy bands.}
\end{deluxetable*}

\subsection{Correlation between Photon Indices and Flux}
\begin{figure*}
\gridline{\fig{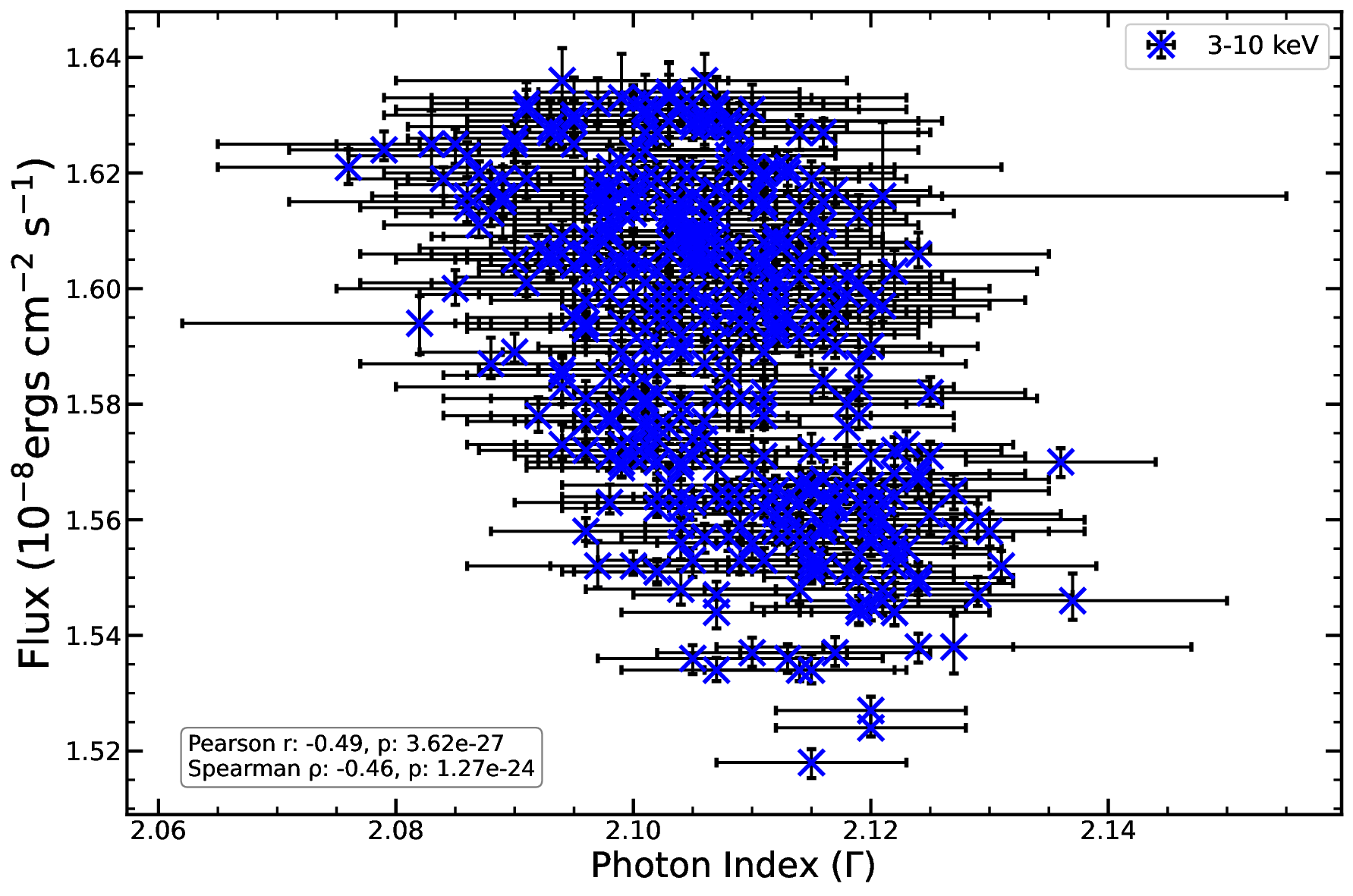}{0.4\textwidth}{(a)}
          \fig{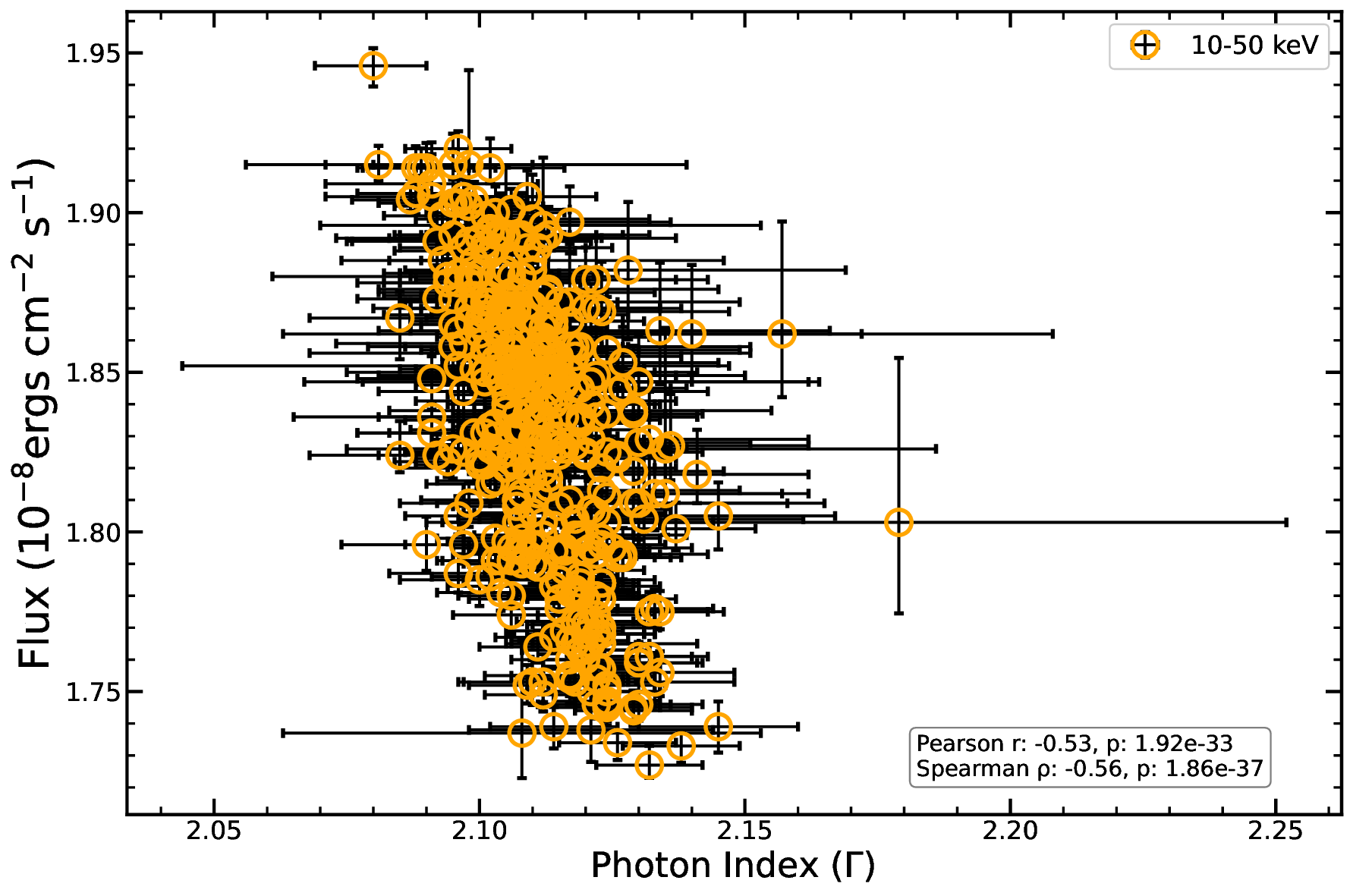}{0.4\textwidth}{(b)}
          }
\gridline{\fig{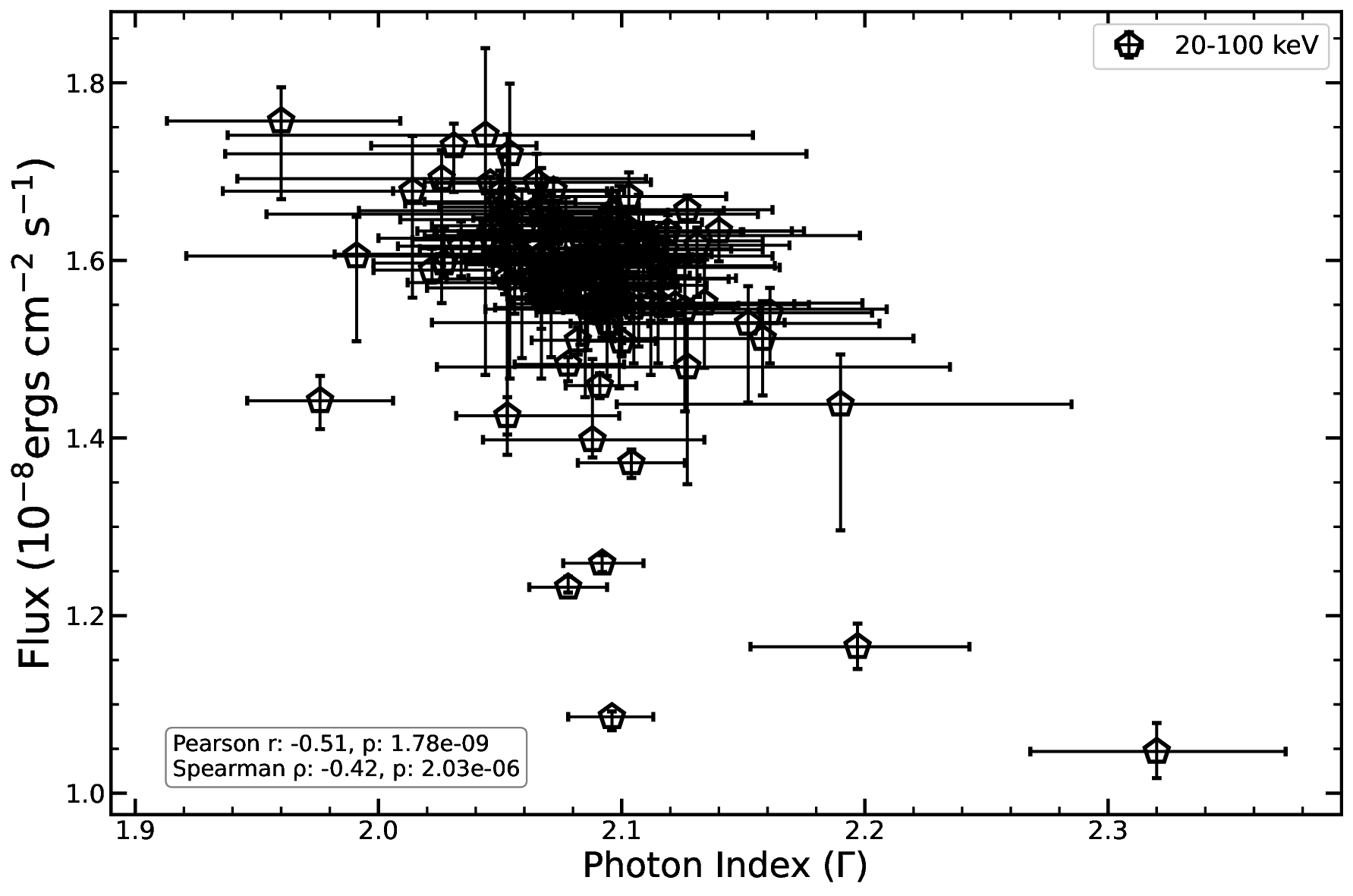}{0.4\textwidth}{(c)}
          }
\gridline{\fig{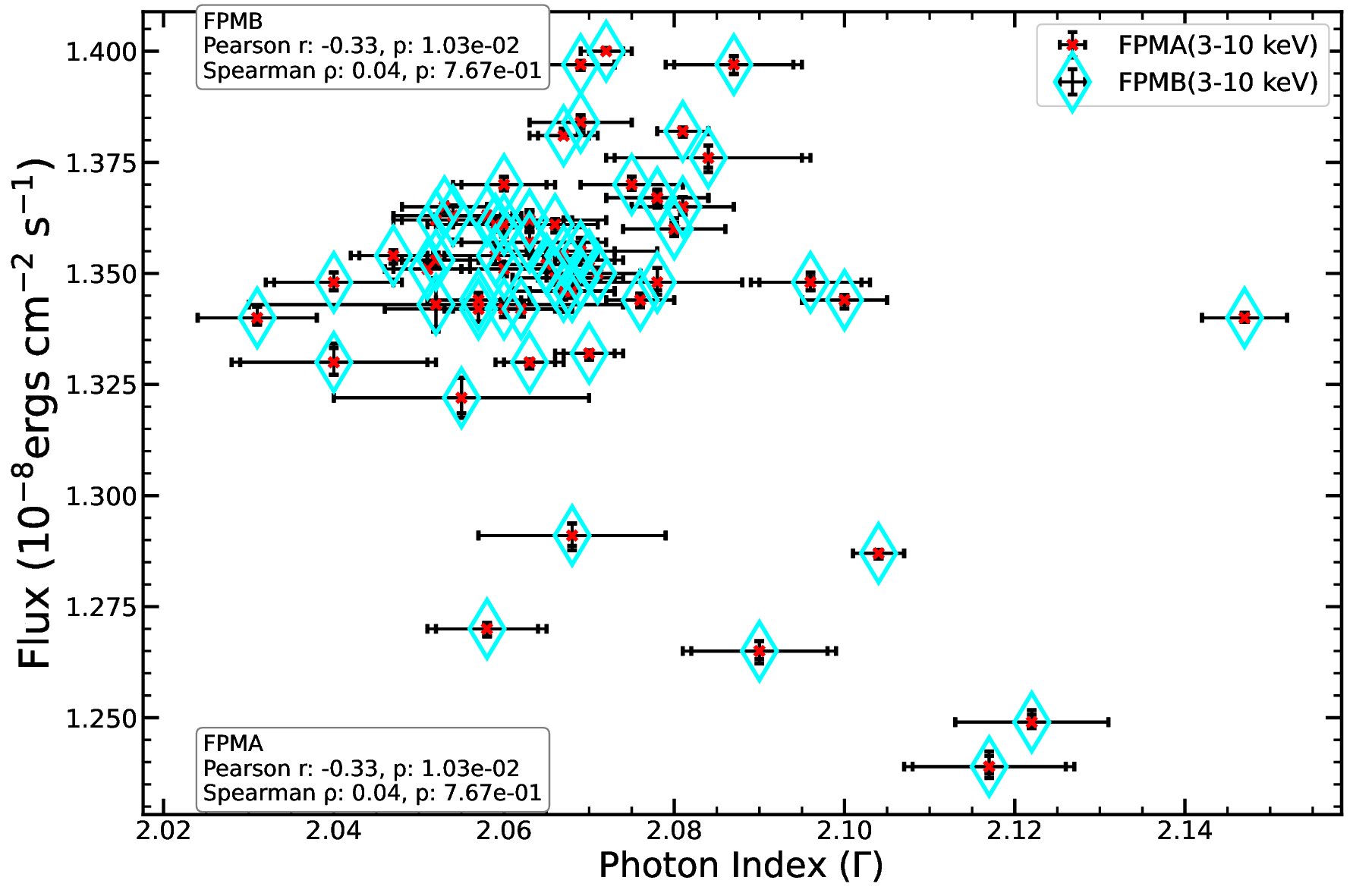}{0.4\textwidth}{(d)}
          \fig{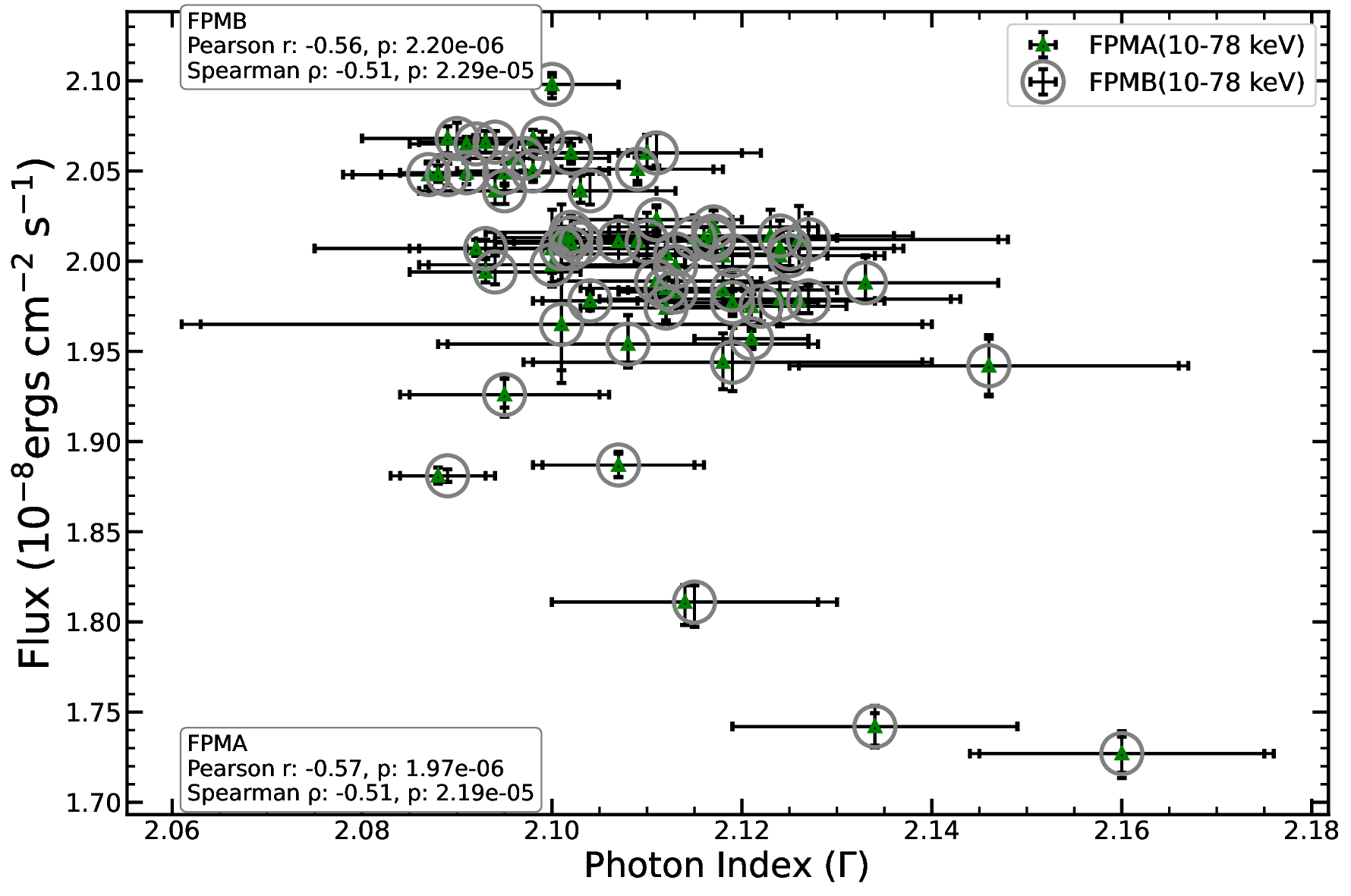}{0.4\textwidth}{(e)}
           }
\caption{Scatter plots of $\Gamma$ vs. Flux: (a) {\it RXTE}/PCA (3--10 keV), (b) {\it RXTE}/PCA (10--50 keV), (c) {\it RXTE}/HEXTE (20--100 keV), (d) {\it NuSTAR}/ FPMA and FPMB (3--10 keV), and (e) {\it NuSTAR}/ FPMA and FPMB (10--78 keV).
{\tt Note:} The correlation coefficients and their corresponding p-values are given in the inset of the figures.
\label{fig:corr}}
\end{figure*}

To find whether there are any correlations between the $\Gamma$ and unabsorbed PL-flux, we scatter-plotted the $\Gamma$ vs. flux for these energy bands and for different instruments (see, Fig~\ref{fig:corr}). A clear correlation is noticed for all the energy bands. To estimate the strength of the correlation, we evaluate Pearson and Spearman correlation coefficients for these plots. The coefficient values are given in Table~\ref{tab:corr}. All the energy bands show a negative Pearson correlation between the $\Gamma$ and PL-flux. Except for the 3--10~keV {\it NuSTAR} observations, other energy bands also show negative Spearman correlation coefficients. We also notice that the correlation coefficients are slightly stronger in high energy bands (10--50~keV or 10--78~keV) compared to the low energy band (3--10~keV). 

\section{Discussions}\label{sec:discus}

We extended a previous study reported in \citet{Augustine&Chang2024} by covering a broader energy range with multiple instruments.
Our analysis of the Crab Nebula's X-ray spectrum using {\it RXTE}/PCA, {\it RXTE}/HEXTE, and {\it NuSTAR} data has provided new insights into its spectral behavior across different energy bands. Below, we discuss the key findings and their implications.

\subsection{Spectral Photon Index and Flux Variation}

The photon index ($\Gamma$) remains around $\sim$ 2.1 across all energy bands studied in this work, with minor variations. However, a systematic difference is observed between the 3--10 keV and 10--50 keV bands, where $\Gamma$ is marginally higher in the latter. This trend is also evident in {\it NuSTAR} observations, with the 3--10 keV $\Gamma$ values being lower than those in the 10--78 keV range. This is consistent with all the earlier observations. The photon index of the Crab's spectrum increases with energy. It starts to decrease (becomes harder) at about 1 GeV, above which another spectral component is observed 
(see, for example, Figure 8 in \citet{Kuiper2001}). The variation of flux shows a $\sim$3 year time-scale in the {\it RXTE}/PCA observation as noticed by previous studies (\citet{Augustine&Chang2024}, see also, Fig.~5 of \citet{Wilson_Hodge2011}). However, the {\it NuSTAR} observations do not show such time-scale in the flux.

To compare the flux between RXTE/PCA and NuSTAR, we also estimated the 10--50~keV unabsorbed flux for NuSTAR (FPMA and FPMB). We notice both the 3--10~keV and 10--50~keV flux for PCA is $\sim$ 1.1-1.2 times higher than FPMA and FPMB flux. We also note that the photon index obtained from NuSTAR data is somewhat smaller than that of RXTE/PCA. It might be due to instruments' calibration issues. This does not affect the results presented in this study because we are actually concerned with variation of fluxes and photon indices, not with their absolute values. 

\subsection{Correlation Between Photon Index and Flux}
A key result of this study is the observed anti-correlation between the photon index ($\Gamma$) and the unabsorbed flux across all energy bands. Both Pearson and Spearman correlation coefficients confirm this trend, with stronger correlations at higher energies (10--50 keV and 10--78 keV) compared to the softer 3--10 keV band. Such an anti-correlation could be related to synchrotron emission, where variations in the nebular magnetic field strength modulate both the flux and spectral slope. This result is consistent with previous findings from Swift BAT Hard X-ray Survey data \citep{Augustine&Chang2024}, further reinforcing the interpretation that magnetic field variations play a crucial role in shaping the X-ray emission. 

\subsection{Implications on Particle Acceleration}
 The negative correlation between $\Gamma$ and flux indicates that higher flux states correspond to harder spectra, implying more efficient acceleration of high-energy electrons during such periods. This behavior may be attributed to localized magnetic reconnection events \citep{Petri2012} or shock acceleration processes near the pulsar wind termination shock \citep{Buehler2012,Cerutti2012,Komissarov2013}.

The observed anti-correlation between photon index and X-ray flux in the Crab Nebula can be interpreted within the framework of diffusive shock acceleration (DSA). Particle-in-cell simulations and analytic models have shown that changes in scattering conditions at the termination shock can significantly alter the electron energy distribution. In particular, anisotropic small-angle scattering increases energy gains and reduces escape probabilities, leading to a flatter electron spectrum and harder X-ray photon index (1 $<\Gamma<$ 2) during high-flux states \citep{Arad2021}. Similarly, variations in the level of magnetic turbulence and shock geometry can modulate the acceleration efficiency: stronger turbulence or higher magnetization can enhance confinement and acceleration, increasing the synchrotron flux while lowering $\Gamma$ \citep{Giacinti2018, Cerutti2020}. These effects naturally produce the `harder-when-brighter' behavior observed by {\it RXTE}, {\it NuSTAR}, (this work) and {\it Swift}/BAT \citep{Augustine&Chang2024}.


The “harder-when-brighter” behavior is also seen in impulsive flares from the Crab Nebula, and has been attributed to magnetic reconnection events that accelerate particles to ultra-relativistic energies on short (hour-to-day) timescales (e.g., \citet{Cerutti2012}). In PWNe like the Crab Nebula, magnetic reconnection occurs through two primary mechanisms: (1) dissipation of striped wind structures formed by alternating magnetic polarities in the pulsar wind, and (2) plasmoid-dominated reconnection at the termination shock front. During active reconnection phases, particles are accelerated to ultra-relativistic energies ($\gamma \sim 10^9$) in current sheets and plasmoid mergers, generating a hard synchrotron spectrum (low $\Gamma$) through enhanced high-energy particle populations \citep{Lyutikov2003,Cerutti2012}. As reconnection subsides, rapid radiative cooling on timescales of minutes to hours depletes these high-energy electrons, steepening the spectral slope ($\Gamma$ increases) while reducing X-ray flux \citep{Kirk2003}. 

Recent particle-in-cell (PIC) simulations provide critical theoretical validation. These studies demonstrate that reconnection produces anisotropic particle distributions with fan-beam geometries perpendicular to current sheets \citep{Cerutti2012}, explaining both the high polarization fractions and rapid variability observed during Crab flares. The simulations further reveal that reconnection acceleration dominates over classical Fermi processes at the termination shock, particularly for generating the highest-energy particles responsible for X-ray emission \citep{Lyutikov2003}.

However, in our study, the anti-correlation persists over year-scale timescales, suggesting a more gradual modulation of the acceleration environment. This long-term behavior may reflect slow evolution of the magnetic field configuration or shock compression at the termination front, which can influence both the spectral index and synchrotron flux over extended periods. Therefore, while reconnection may explain rapid flaring events, the year-scale spectral trends observed here are more consistent with large-scale, secular changes in the nebular conditions.

\begin{acknowledgments}
We would like to thank the anonymous referee for the constructive comments and suggestions, which have signiﬁcantly improved the quality of this paper.
D.C., H.K.C, K.A.A. and T.H.L acknowledge the grants NSPO-P-109221 of the Taiwan Space Agency (TASA) and NSTC-112-2112-M-007-053 of the National Science and Technology Council (NSTC) of Taiwan. D.D. acknowledges the visiting research grant of National Tsing Hua University, Taiwan, with the support from NSTC (NSTC 113-2811-M-007-010).
\end{acknowledgments}





%
\vspace{5mm}
\facilities{{\it RXTE}(PCA and HEXTE), {\it NuSTAR}(FPMA and FPMB)}
\software{ 
          HeaSoft \url{https://heasarc.gsfc.nasa.gov/docs/software/heasoft/} \\ 
          FTOOLs \url{https://heasarc.gsfc.nasa.gov/ftools}, \citep{2014ascl.soft08004N}
          }




\bibliography{sample7}{}
\bibliographystyle{aasjournalv7}



\end{document}